\documentclass[%
 aip,
 jmp,%
 amsmath,amssymb,
preprint,%
]{revtex4-1}

\usepackage{graphicx}
\usepackage{dcolumn}
\usepackage{bm}

\begin{document}

\preprint{AIP/123-QED}

\title[]{Phase diagram and piezoelectric response of (Ba$_{1-x}$Ca$_x$)(Zr$_{0.1}$Ti$_{0.9}$)O$_3$ solid solution}

\author{Desheng Fu}
 \email{ddsfu@ipc.shizuoka.ac.jp}
 \homepage{http://www.grl.shizuoka.ac.jp/~ddsfu/}
 \affiliation{%
  Department of Electronics and Materials Science, Graduate School of Engineering, Shizuoka University,3-5-1
  Johoku, Naka-ku, Hamamatsu 432-8561, Japan
 }%

\author{Yuto  Kamai}
\author{Naonori Sakamoto}
\author{Naoki Wakiya}
\author{Hisao Suzuki}
 \affiliation{%
 Department of Electronics and Materials Science, Graduate School of Engineering, Shizuoka University,3-5-1
 Johoku, Naka-ku, Hamamatsu 432-8561, Japan
  }%

\author{Mitsuru Itoh}%
\affiliation{%
Materials and Structures Laboratory, Tokyo Institute of Technology,
4259 Nagatsuta, Yokohama 226-8503,
Japan
}%

\date{\today}

\begin{abstract}
We report the phase diagram of
(Ba$_{1-x}$Ca$_x$)(Zr$_{0.1}$Ti$_{0.9}$)O$_3$ solid solution. It is
found that substitution of smaller Ca ions for Ba ions  can slightly
increase the cubic-tetragonal($T$) para-ferroelectric phase
transition temperature and strongly decrease the $T$-orthorhombic
($O$) and $O$-rhombohedral ($R$) transition. This unique
ferroelectric phase evolution is attributed to  Ca off-centering
effects. More importantly, lowering of the  $T-O$ or $O-R$ phase
transitions allows us to prepare the piezoelectric ceramics with a
strain response as high as $S/E\approx800$ pm/V ($E$=10 kV/cm) over
a wide range of compositions with $x\approx 0.1 - 0. 18$ at room
temperature, which may be interesting for piezoelectric
applications.

\end{abstract}

\maketitle

BaTiO$_3$(BTO) was the first commercial piezoelectric ceramic
\cite{Jaffe}, and it is being used in industry, for example, in
sonar transducers used in fishfinders.\cite{Honda} Its large
piezoelectric response is essentially due to  large lattice
distortion under an electric field \cite{Tazaki} and the easy
switching of ferroelectric domains facilitated by a low coercive
field.\cite{Wieder} Due to these large piezoelectric effects, BTO is
also commonly used to develop new lead-free piezoelectric
materials.\cite{Rodel,Fu2008}

The ferroelectricity of BTO with an ABO$_3$ perovskite structure
originates from the Ti-shift  in the BO$_6$ octahedron, giving rise
to three ferroelectric phases in the crystal: the tetragonal($T$)
phase with spontaneous polarization $P_{\rm s}\parallel[001]_{\rm
c}$, orthorhombic($O$) phase with $P_{\rm s}\parallel[011]_{\rm c}$
and rhombohedral($R$) phase  with $P_{\rm s}\parallel[111]_{\rm
c}$.\cite{Hippel} Either A- or B-site substitution can change the
state of spontaneous polarization, providing a variety of modifying
phase evolutions and thus modifying  the physical properties of
BTO.\cite{Jaffe} Among the various substitutions in the  BTO system,
Zr-substitution for B-site ions \cite{Kell,Dobal,Rehrig,Yu} and
Ca-substitution for A-site ions
\cite{Fu2008,Mitsui,FuPRL,FuJPC2010,Shimizu} are of particular
interest.

When substituting Zr  for Ti in BTO, one can observe an abundance of
phase evolutions spanning from ferroelectric phases to relaxor or
polar cluster states.\cite{Maiti}  Even in the region of
ferroelectric phases for Zr-substitution amounts of less than
$\approx20$ mol\%,  ferroelectric phase evolutions are   very
complicated. In contrast to the monotonic decrease in the cubic
($C$)-$T$ phase transition, the $T$-$O$ and $O$-$R$ phase
transitions initially increase, reach a maximum at approximately 10
mol\% substitution, and then decrease with increasing the
Zr-substitution amount.\cite{Kell,LB} Around the composition
corresponding to the maximum transition point, the $T-O$ and $O-R$
phase transitions approach the $C-T$ para-ferroelectric transition
point. More interestingly, this solid solution shows a very large
piezoelectric response,\cite{Rehrig,Yu} which  has attracted much
attention in the development of  piezoelectric materials.\cite{Liu,
Xue}

When substituting smaller Ca ions for bulky Ba ions, unique phase
evolution\cite{Mitsui,FuPRL} and interesting physical phenomena
\cite{FuPRL,FuJPC2010,Shimizu} have  been observed. In contrast to
the reduction of  the $C-T$ para-ferroelectric transition for Zr
substitution,\cite{Kell,LB} this phase transition remains nearly
unchanged for Ca-substitution in the limit of solid
solution.\cite{Mitsui,FuPRL} However, the $T-O$ and $O-R$
transitions are shifted to lower temperatures such that $T^i\propto
(x^i-x)^{1/2}$, where $i$ denotes the phase boundaries of the $T-O$
and $O-R$ transitions, $x$ is the Ca-substitution amount,
$x^{O-R}=0.180$, and $x^{T-O}=0.233$. Although substitution of Ca
for Ba results in the reduction of unit cell volume, which is
naturally assumed to suppress the stability of the ferroelectric
phase, the  $C-T$ transition remains nearly
unchanged.\cite{Mitsui,FuPRL} This result suggests that a
polarization component other than Ti displacement exists in
(Ba$_{1-x}$Ca$_x$)TiO$_3$ crystal to compensate for pressure-induced
ferroelectric instability and stabilize the ferroelectric $T$ phase.
This assumption is strongly supported by a first-principles
calculation,\cite{FuPRL} which predicts a Ca off-centering shift of
0.1 ${\AA}$ that is larger than the Ti shift in pure BaTiO$_3$ (0.05
${\AA}$).\cite{LB} This off-center displacement by smaller ions thus
provides us an approach to control the ferroelectricity of
perovskite oxide. Here, we examine the effects of Ca off-centering
on the phase evolution and piezoelectric response in
(Ba$_{1-x}$Ca$_x$)(Zr$_{0.1}$Ti$_{0.9})$O$_3$ (BCZT) solid solution.

BCZT ceramics were prepared by a solid-state reaction approach.
Mixtures of BaCO$_3$, CaCO$_3$, ZrO$_2$ and TiO$_2$ were calcined at
1823 K for 3 h. The calcined powders were ground, pressed and
sintered at 1823 K for 5 h. The ceramic pellets were then
electroplated with silver for electrical measurements. X-ray power
diffraction patterns were obtained using  a Bruker AXS D8 ADVANCE
diffractometer.  The permittivities were measured using a
Hewlett-Packard Precision LCR meter (HP4284A) at an ac level of
1V/mm.  A cryogenic temperature system (Niki Glass LTS-250-TL-4W)
was used to control the temperature within the range of 4 - 450 K.
The electric-field-induced strain and dielectric hysteresis loops
were measured using a ferroelectric measurement system (Toyo
Corporation FCE-3) equipped with an Iwatsu ST-3541 capacitive
displacement meter having a linearity of 0.1 \% and  a resolution of
0.3 nm.

\begin{figure}
\includegraphics[width=12cm]{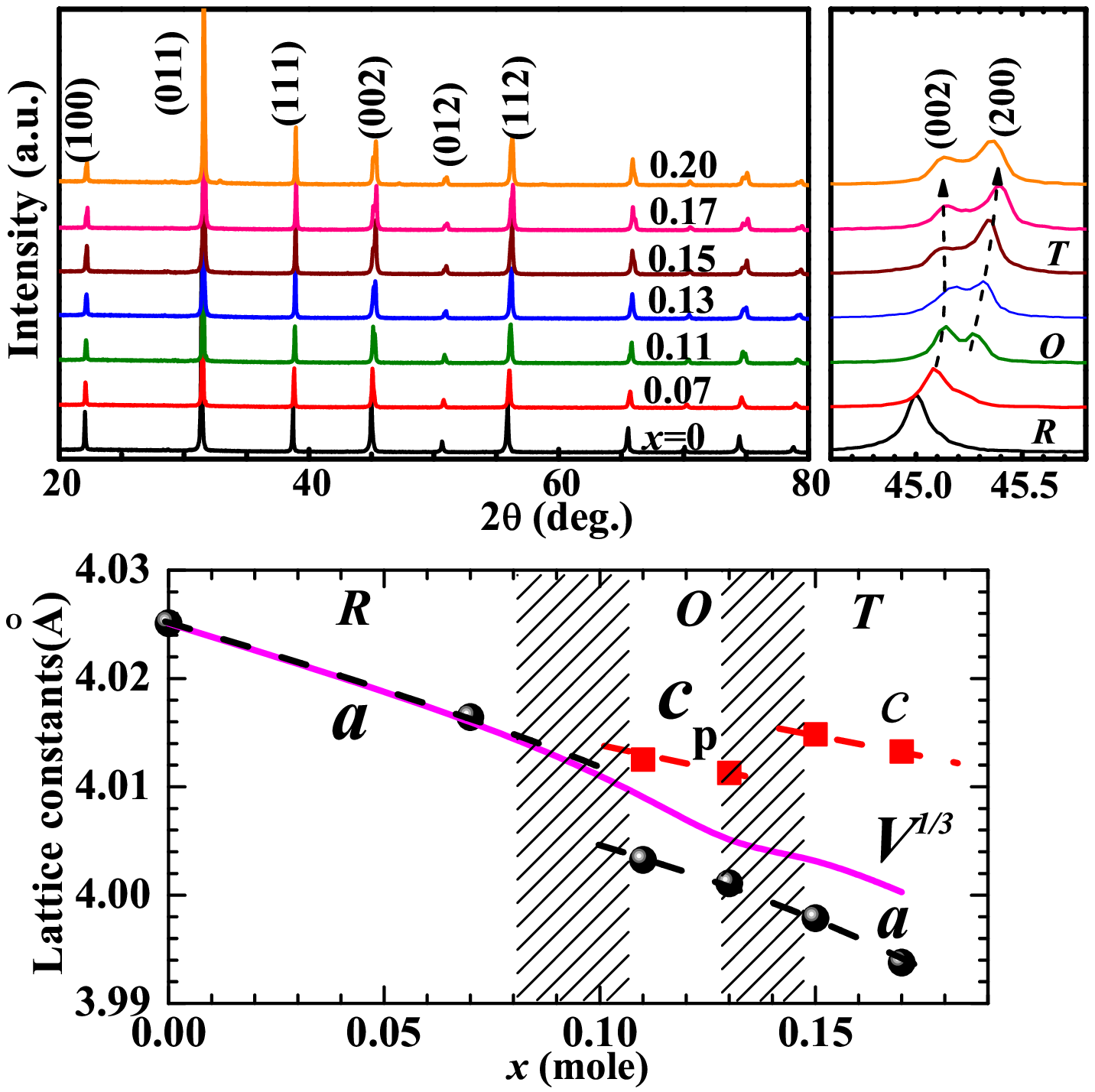}
\caption{\label{Fig1} X-ray diffraction patterns  and  unit cell
parameters of BCZT solid solutions obtained at room temperature.}
\end{figure}

Figure \ref{Fig1} shows X-ray diffraction patterns of BCZT solid
solutions obtained at room temperature, from which we can observe
the structure changing with composition. Very weak diffractions,
which can be indexed by the orthorhombic structure of CaTiO$_3$,
were found in the compound with $x=0.20$. This suggests that the
limit of solid solution is less than 20 mol\% in this system. Within
the limit of solid solution,  X-ray diffraction clearly indicated
that there were structural changes in BCZT from $R$ phase to $O$
phase and then to $T$ phase with increasing Zr-substitution amounts.
X-ray diffraction patterns were well indexed as rhombohedral
symmetry for  samples with $x\leq0.7$, as orthorhombic symmetry for
sample with $x=0.11 $, and as tetragonal symmetry for samples with
$x\geq 0.13$. The reflections in the 2$\theta$ range of 60 degree to
120 degree were then used to calculate the unit cell parameters by
the method of least squares.  The standard deviations of unit cell
parameters were less than 0.0008 ${\AA}$ for all calculations.  As
shown by the variation of the lattice constant, due to the small
ionic radius of Ca ions relative to Ba ions, substitution of Ca for
Ba leads to a decrease in unit cell volume. It was  found that the
ferroelectric lattice distortion was very small in the BCZT solid
solution. The distortion angles $\alpha$ in the $R$ phase and
$\beta$ of the monoclinic unit cell in the  $O$ phase  had only
0.01$^\circ$ and 0.1 $^\circ$ deviations from the right angle,
respectively, while the tetragonality $c/a$ has a value of 1.005 for
$x\geq0.15$ in the $T$ phase.

\begin{figure}
\includegraphics[width=10cm]{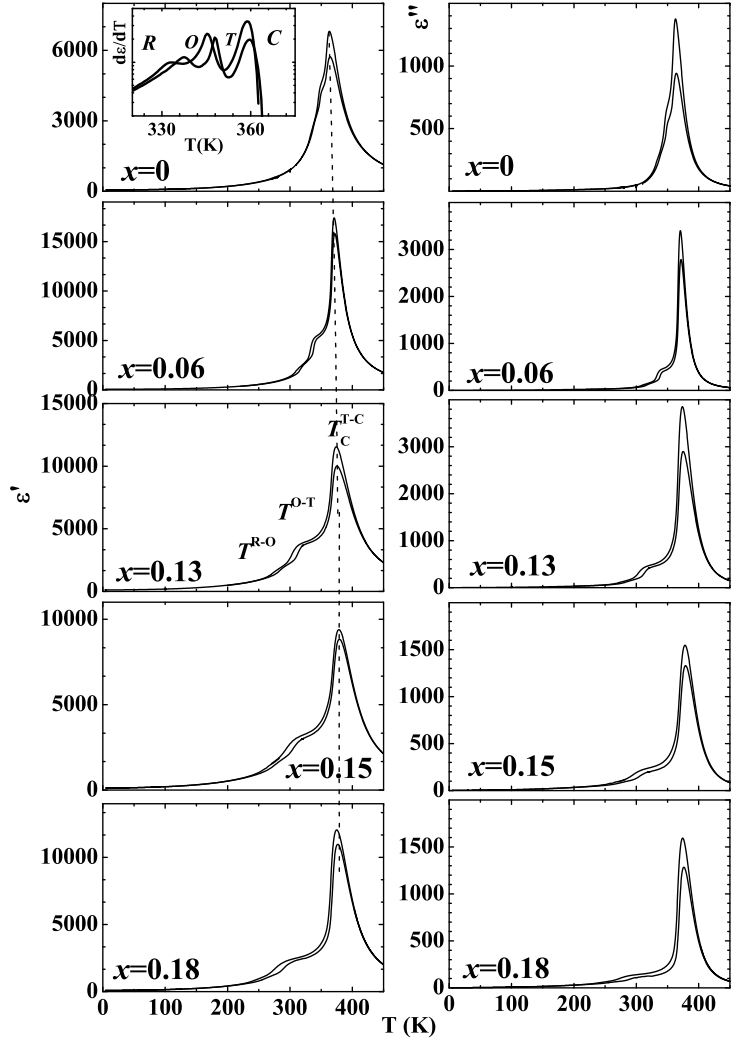}
\caption{\label{Fig2} Temperature variation of the real
($\varepsilon'$) and imaginary ($\varepsilon "$) parts of
permittivity in  BCZT solid solutions. Data were obtained at 100 kHz
and an ac level of 1 V/mm.}
\end{figure}

\begin{figure}
\includegraphics[width=10cm]{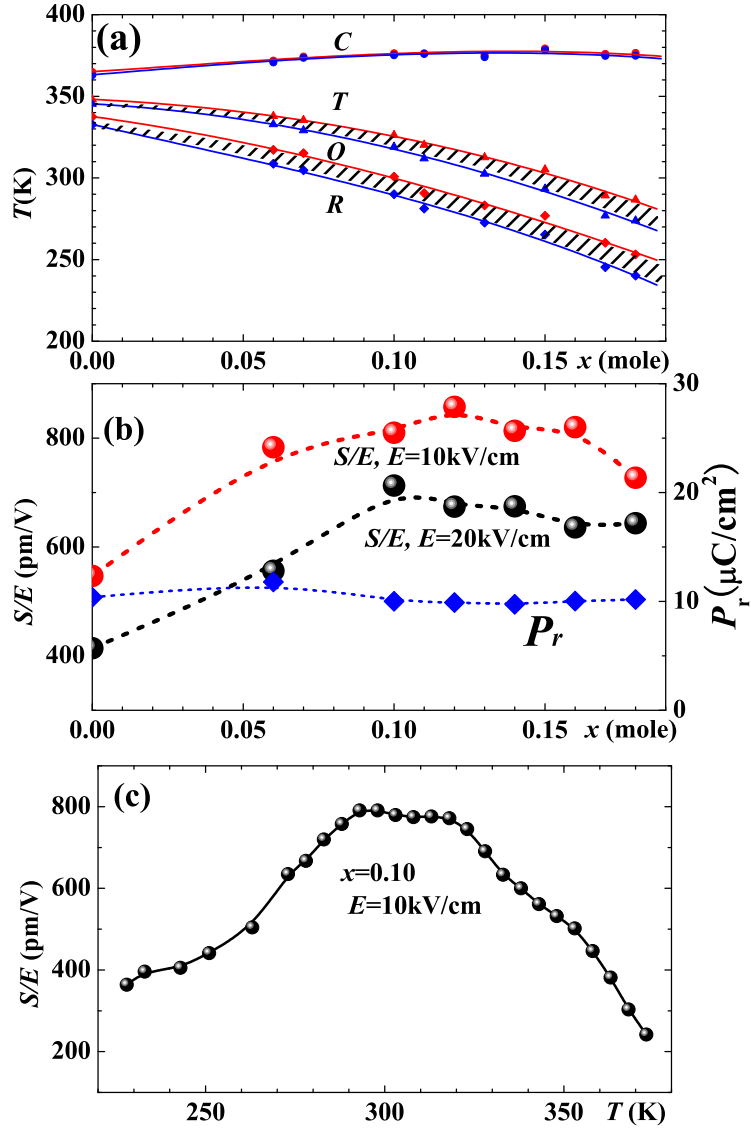}
\caption{\label{Fig3} (a) Phase diagram of BCZT. The slanted line
region indicates the thermal hysteresis of the phase transition.
Thermal hysteresis of the $C-T$ transition is very small and is less
than 2 K. (b) Variation of remanent polarization $P_{\rm r}$ and
strain level $S$ at $E$=10 and 20 kV/cm  with composition at $T$=295
K. (c) Temperature dependence of strain level obtained at $E$=10
kV/cm for $x=0.1$.}
\end{figure}

To observe the phase evolution with temperature, we measured the
temperature variation of the permittivity. The results are shown in
Fig. \ref{Fig2}. For a composition of $x=0$, the para-ferroelectric
phase transition occurred at 363 K. Since the $T-O$ and $O-R$ phase
transitions are very close to the $C-T$ transition, the transition
temperatures are not easy to determine directly from the
permittivity curve. However, these two transitions can be clearly
determined from the differential curve, as shown in the inset of
Fig.\ref{Fig2}. Here, we use the maximum point of the differential
curve to define the $T-O$ and $O-R$ phase transitions; this
contrasts with  the case of the $C-T$ transition, which can be
determined directly from the maximum of the permittivity curve.
Using this approach, we established the phase diagram for this
system, which is shown  in Fig. \ref{Fig3}. Similarly to the case of
(Ba$_{1-x}$Ca$_x$)TiO$_3$,\cite{Mitsui,FuPRL} the $C-T$
para-ferroelectric phase transition does not decrease with the
reduction in unit cell volume associated with an increase in the
Ca-substitution amount. A slight difference compared with
(Ba$_{1-x}$Ca$_x$)TiO$_3$  is that the $C-T$ transition in BCZT
initially increases from 363 K for $x=0$ to 376 K for $x=0.1$, after
which it seems to reach  saturation with further substitution. The
change in the $T-O$ and $O-R$ phase transitions in BCZT is also very
similar to that in (Ba$_{1-x}$Ca$_x$)TiO$_3$.\cite{Mitsui,FuPRL}
Both the $T-O$ and $O-R$ phase transitions decrease with
Ca-substitution. The similarity of the ferroelectric phase evolution
in BCZT with that in (Ba$_{1-x}$Ca$_x$)TiO$_3$ strongly suggests
that Ca off-centering effects play a critical role in tuning the
polarization states in these two systems.\cite{FuPRL}

\begin{figure}
\includegraphics[width=12cm]{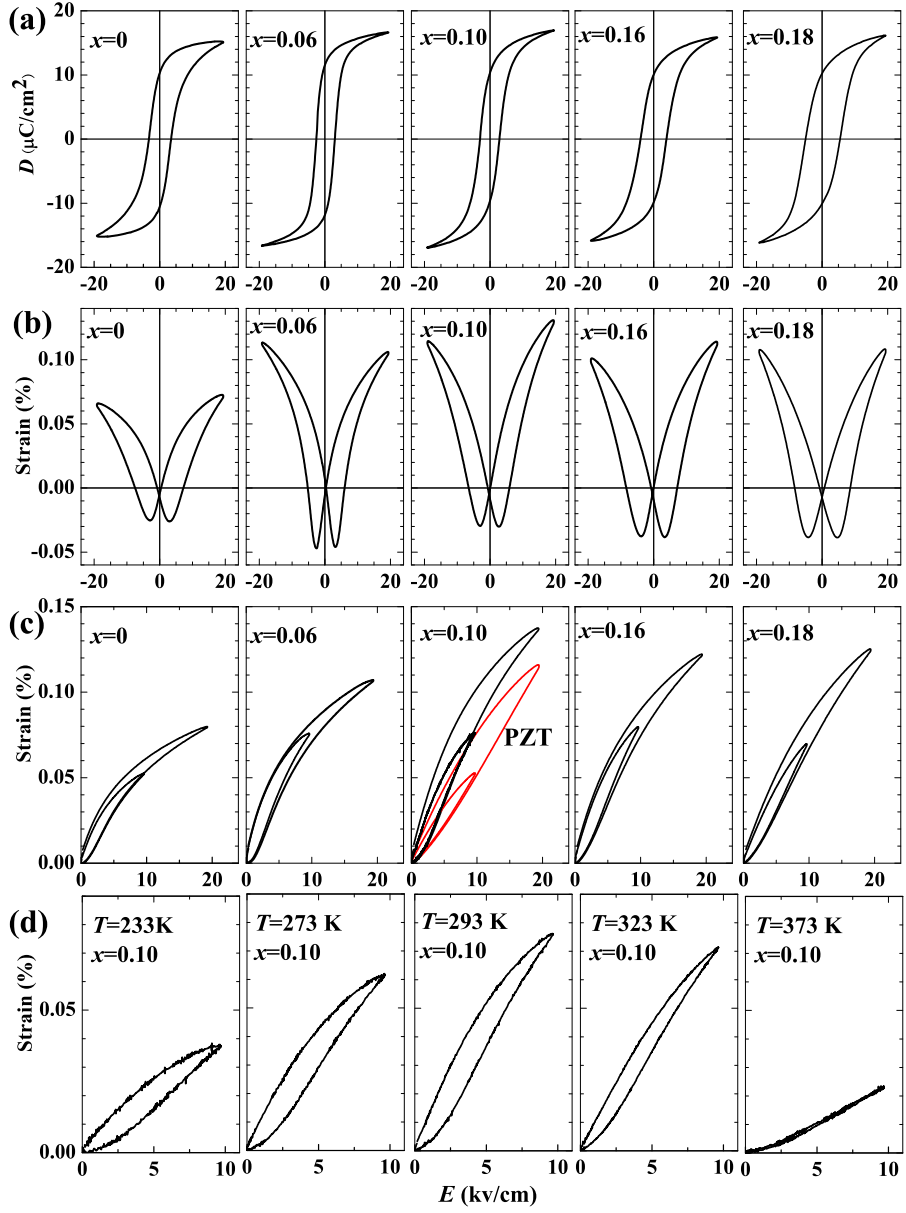}
\caption{\label{Fig4} $D-E$  hysteresis loop (a), and strain induced
by the bipolar (b) and unipolar (c) electric fields in the BCZT
piezoelectric ceramics at $T$=295 K. For comparison, the strain
response of PZT ceramic in an unipolar field is also shown. (d)
shows the temperature dependence of strain induced by the unipolar
electric field $E$=10 kV/cm for $x=0.1$.}
\end{figure}

We also examined the ferroelectric and piezoelectric behaviors in
this system  and show the results in Fig. \ref{Fig4}. The remanent
polarization $P_r$ has an approximate value of 10 $\mu$C/cm$^2$ and
seems to remain unchanged within the limit of solid solution
(Fig.\ref{Fig3}(b)), which is  very similar to what is observed in
(Ba$_{1-x}$Ca$_x$)TiO$_3$.\cite{FuJPC2010} The coercive field of a
BCZT ceramic usually falls in the range of 3 - 5 kV/cm. This low
coercive field is favorable for electrical  poling in a ceramic
sample.

More interestingly, BCZT ceramics are extremely soft elastically. A
surprisingly high elastic deformation of $\approx0.15$\% was
observed in the ceramic under the application of a bipolar electric
field of 20 kV/cm. This large electromechanical coupling was further
confirmed by the strain response in a unipolar field. From the
results of unipolar field measurements,  it is clear that a strain
level higher than that of commercial PZT ceramic is present in  BCZT
ceramics at the same electric field. To understand the large strain
response in BCZT ceramics , we analyzed the variation of the
effective piezoelectric constants defined by $S/E$ (=the ratio of
maximum strain to the applied electric field) and compared this
variation with the phase evolution in the phase diagram. From Fig.
\ref{Fig3}(b), it is immediately clear that at a measurement
temperature of $T$=295 K, BCZT solid solutions undergo phase changes
between $R-O$ or $O-T$ states over compositions ranging from
$x\approx$0.1 to $x\approx$0.18. This phase evolution is accompanied
by the appearance of large effective piezoelectric constants for
BCZT. Values of $S/E$ greater than 800 pm/V were observed in this
wide composition region. It is reasonable to consider that
structural transformation plays a critical role in the large
piezoelectric response in these solid solutions. As observed in the
X-ray diffraction measurements, the ferroelectric distortions among
the $T$, $O$ and $R$ ferroelectric phase are very small. Moreover,
the difference in transition temperature between $R-O$ and $O-T$
transitions is approximately 20 K, indicating that the energy
difference between these ferroelectric phases is very small. The
small ferroelectric distortion together with the small energy
difference between these ferroelectric phases are thought to
facilitate the phase transformations, resulting in the large strain
response under the application of an electric field in these solid
solutions.

We further measured the dependence of strain response on the
temperature for ceramic sample with $x=0.1$ having large
piezoelectric effects around room temperature in the temperature
range of 228 K - 373 K. The results are summarized in
Fig.\ref{Fig4}(d) and Fig.\ref{Fig3}(c). In this composition, the
$R-O$ and $O-T$ phase boundaries cover the temperature range of 290
K - 320 K. Just within this temperature range, the ceramic sample
shows very large elastic deformation under the application of
electric field. As temperature departs from this temperature range,
the strain level also decreases. This again suggests that the $R-O$
and $O-T$ phase boundaries are responsible for the large
electromechanical coupling effects in BCZT solid solutions. An
interesting phenomenon for this material is that the strain response
is very stable in the temperature range mentioned above, which might
be of significance for the practical applications.

Here, it should be mentioned a recent report by Liu  and Ren
\cite{Liu} on system of $(1-y)$Ba
(Ti$_{0.8}$Zr$_{0.2}$)O$_3$-$y$Ba$_{0.7}$Ca$_{0.3}$TiO$_3$, which
shows a  high piezoelectric coefficient of $d_{33}$=620 pC/N at
$y=0.5$,  triggering a great interest in this composition in the
last two years. Liu and Ren state that the high piezoelectricity of
composition is due to a morphortropic phase boundary (MPB) between a
ferroelectric $R$ and a ferroelectric $T$ phases, which leads to a
nearly vanishing polarization anisotropy and thus facilitates
polarization rotation between $[001]_c T$ and $[111]_c R$ states.
Actually, this composition  is exactly the same as that demonstrated
in the present system of
(Ba$_{1-x}$Ca$_x$)(Zr$_{0.1}$Ti$_{0.9}$)O$_3$ with $x=0.15$. In
contrast to Liu and Ren's statement, our phase diagram dose not
support the presence of a direct phase transition between $R$ and
$T$ phase in this composition. Instead, our results indicate that
this  composition has the successive phase transitions of
 $C-T-O-R$ as  BTO. This conclusion is also supported by the pyroelectric current measurements reported by Benabdallah {\it et al.}\cite{Benabdallah},
in which three sharp peaks of pyroelectric currents corresponding to
three phase transitions are demonstrated. Actually, when carefully
examining the temperature dependence of permittivity of Liu and
Ren's sample with $y=0.5$, one can find that in addition to  the
$T-C$ transition and the  so-called  "$R-T$" transition at $T_{\rm
R-T}$ there is another phase transition at lower
temperature.\cite{Liu,Xue} Therefore, we consider that the
polymorphic phase transitions observed in  our study can offer a
reasonable explanation for the large piezoelectric response in the
solid solution of (Ba$_{1-x}$Ca$_x$)(Zr$_{0.1}$Ti$_{0.9}$)O$_3$ with
$x=0.15$. Actually, our samples with $x=0.1-0.18 $ having $R-O$ or
$O-T$ phase transformations around room temperature all show very
large piezoelectric response comparable to that with $x=0.15$.

In summary, we have obtained the phase diagram of
(Ba$_{1-x}$Ca$_x$)(Zr$_{0.1}$Ti$_{0.9}$)O$_3$ solid solution. It was
demonstrated that Ca off-centering plays a critical role in
stabilizing the ferroelectric phase and tuning the polarization
state in the (Ba$_{1-x}$Ca$_x$)(Zr$_{0.1}$Ti$_{0.9}$)O$_3$ system.
The Ca off-centering effect also allowed us to shift the $R-O$ and
$O-T$ phase boundaries to room temperature, leading  to the
occurrence of electromechanical coupling effects, comparable to
those of  PZT, in the  system over a large composition range.

\begin{acknowledgments}
This work was partly supported by Grants-in-Aid for Scientific
Research from The Ministry of Education, Culture, Sports, Science
and Technology of Japan
\end{acknowledgments}


\clearpage


\clearpage \clearpage

\clearpage

\clearpage

\clearpage



\end{document}